\documentclass[notoc]{tufte-handout}

\usepackage{amssymb,amsmath,amsthm}
\usepackage{graphicx,ctable,booktabs}
\usepackage{listings, verbatim, url}
\usepackage{setspace}

\titlecontents{section}[1.8pc]
  {\addvspace{10pt}\itshape \large}
  {\contentslabel[\thecontentslabel]{1.8pc}}
  {}
  {\quad\thecontentspage}

\addtocontents{toc}{\vspace{-0.4cm}}

\makeatletter
\renewcommand{\@seccntformat}[1]{}
\makeatother

\newcommand{\approptoinn}[2]{\mathrel{\vcenter{
  \offinterlineskip\halign{\hfil$##$\cr
    #1\propto\cr\noalign{\kern2pt}#1\sim\cr\noalign{\kern-2pt}}}}}

\title{Coconuts and Islanders: A Statistics-First Guide to the Boltzmann Distribution}
\author{Brian Zhang, University of Oxford}
\date{\thedate}

\begin{document}

\begin{onehalfspacing}
\maketitle
\end{onehalfspacing}

\thispagestyle{empty}


\section*{Abstract}




\noindent The Boltzmann distribution is one of the key equations of thermal physics and is widely used in machine learning as well. Here I derive a Boltzmann distribution in a simple pedagogical example using only tools from a first-year probability course. The example is called ``coconuts and islanders'' and was taught to me by my father, Shoucheng Zhang (1963 - 2018), to whom these notes are dedicated. By focusing on this simple story, which can be easily simulated on a computer, I aim to provide a more accessible and intuitive presentation of the Boltzmann distribution. Yet I hope this exposition also inspires deep thinking about statistical physics. For instance, I show that the coconuts and islanders story illuminates a connection between the ``fundamental assumption of statistical mechanics''---all microstates are equally probable---and the statistical property of detailed balance.

\setcounter{secnumdepth}{3}
\setcounter{tocdepth}{1}


\tableofcontents

\pagebreak

\section{Introduction}

The Boltzmann distribution, named after Ludwig Boltzmann who discovered it in the late 19th century, forms the foundation of statistical mechanics and its applications to materials science and chemistry. In the second half of the 20th century, the Boltzmann distribution also made its way into the machine learning community, inspiring a class of so-called ``energy-based'' statistical models. Despite its wide use, I found the Boltzmann distribution to be a bit of a mystery the first several times I encountered it in my studies. In this writeup, I hope to share some ways of thinking about the Boltzmann distribution that I wish I had been exposed to earlier on.


The main part of this exposition centers around a parable my Dad had told me and my sister when we were young. Like many parental sermons, I didn't take his words much to heart at the time. Then in my first year out of college, I---as someone who had majored in physics---was trying to explain the Boltzmann distribution to a friend, and realized how much of it was still opaque to me. A continent away from home, I recalled my Dad's parable and worked out some of my own calculations on a train journey. For over a year, I remained hooked on this problem, finding new extensions to calculate and code up in my free time. Through it all, I came to understand the Boltzmann distribution better than I could have through reading any textbook.

I shared some preliminary versions of this writeup with my Dad, who was both flattered and intrigued: I had taken some different routes of calculation than what he had in mind. These notes sat unfinished, when in December 2018 my Dad sadly passed away. In memory of him, I've chosen to polish my notes and rework them to add some framing and stories around my Dad. In the margin notes, I've added some anecdotes from my Dad's enthusiastic perspectives on the subject. In the conclusion, I share some words about how this simple problem embodies features of my Dad's passions in both teaching and research.

In the rest of this section, I provide some background on the Boltzmann distribution to motivate its broad utility. Then, starting in Section \ref{story}, I dive into the main story.



\subsection{The Boltzmann distribution}
For readers who aren't familiar with it, the Boltzmann distribution takes a system in statistical physics and assigns relative probabilities over the possible outcomes. The main condition is that the system is at a fixed temperature $T$.\footnote{What is temperature, you ask? See the appendix for some notes.}


Assuming that the system can take many different states $s$, each with its own energy $E(s)$, the Boltzmann distribution says that the probability of observing the system in state $s$ is given up to a proportionality constant by
$$
p(s) \propto e^{-E(s) / kT}.
$$
Here $k$ is the Boltzmann constant, and $T$ is measured in Kelvins.


If we assume that all the possible states of the system are indexed by a discrete set $S$, then we can normalize the Boltzmann distribution using the \emph{partition function}:
$$
Z = \sum_{s \in S} e^{-E(s) / kT}
$$
This allows us to write:
$$
p(s) = \frac{1}{Z} e^{-E(s) / kT}.
$$

The Boltzmann distribution can be applied to describe the states of an atom, molecule, ensemble of physical particles, or even some biological systems. 
We give three examples of scenarios that are well-modeled by the Boltzmann distribution.

\subsection{The distribution of molecular speeds in a gas}\label{s:gas}
Today we know that matter is made up of atoms, but until the early 20th century, the atomic principle was a contested hypothesis. Yet in 1738, Daniel Bernoulli gave an explanation of properties of a gas by assuming a gas was made up of many randomly moving molecules.\footnote{The publication, \emph{Hydrodynamica}, also contained a description of Bernoulli's principle for fluids.} Bernoulli's model, called the kinetic theory of gases, explains that when the volume of a gas is shrunk, the pressure will increase because molecules hit the boundaries more frequently. Bernoulli also postulated that heating the gas speeds up the motion of molecules, also increasing the pressure. These relationships form the basis of the ideal gas law, $PV = nRT$.

Bernoulli's model can be used to derive a direct relationship between temperature $T$ and the mean squared molecular speed $v^2$. The Maxwell-Boltzmann distribution, formulated in the 1860s-70s, goes further by describing the entire distribution of molecular speeds within the gas. The derivation is straightforward given the Boltzmann distribution, once one identifies the energy of a single gas molecule as $E = \frac{1}{2}mv^2$, the formula for kinetic energy. For a given speed $v$, the probability of observing that speed is proportional to\footnote{We include a factor of $v^2$ with the Boltzmann factor because the possible velocity vectors with speed between $v$ and $v + dv$ form a spherical shell of size $4\pi v^2 dv$.}
$$
p(v) \propto v^2 \exp \left( - \frac{E(v)}{kT} \right) = v^2 \exp \left( - \frac{mv^2}{2kT} \right).
$$

At low temperatures, quantum mechanical effects kick in which modify the Boltzmann distribution, and consequently the distribution of molecular speeds. However, at room temperature or higher, the Maxwell-Boltzmann distribution is usually quite an accurate approximation, as has been investigated experimentally.\cite{Thallium} Such experiments typically construct a heated oven of gas, allow molecules to escape through a small slit, and measure the velocities upon escape.


\subsection{Protein folding}\label{s:protein}
In cells, proteins are created by ribosomes which chain together amino acids. The ribosomes construct this chain by reading off a ``program'' from an RNA molecule: each triplet of RNA letters specifies which of 22 amino acids to append next, until a ``stop'' instruction is reached. This protein chain, initially like a linear piece of string, quickly coils up into a stable configuration. The resulting 3D shape of the folded protein is essential to its function in the cell, such that one of the grand challenges of computational biology is to predict a protein's folded configuration from its amino acid sequence.

The folded configuration of a protein is the one that minimizes its energy -- a sum over the energies from many molecular interactions, such as hydrogen bonds and van der Waals forces. This is because according to the Boltzmann distribution, among all different states, the minimum-energy state is most probable. Assume a simplistic case where a folded protein has energy $0$, while there are $N$ unfolded states each with energy $E$. Then the partition function is given by:
$$
Z = \sum_{s\in S}e^{-E(s)/kT} = 1 + Ne^{-E/kT}.
$$
So the probability of the folded state is:
$$
p(\mbox{folded}) = \frac{1}{Z} e^{0/kT} = \frac{1}{1 + Ne^{-E/kT}} = \frac{e^{E/kT}}{e^{E/kT} + N}.
$$
If the energy gap $E$ is sufficiently large, the probability of the folded state will overwhelm all other states, even though these are more numerous by a factor of $N$ to $1$.

However, we can also investigate the dependence of the above on temperature. If we increase the temperature $T$, then the quantity $e^{E/kT}$ begins to shrink, and once it goes below $N$, the folded state is no longer overwhelmingly probable.\footnote{If we analyze the condition for unfolding, setting a threshold of $p(\mbox{folded}) < \frac{1}{2}$, the condition becomes $e^{E/kT} < N$, or $E < kT \ln N$. This can be rearranged as $E - TS < 0$, where $S = k \ln N$ is the entropy. In chemistry, one has the analogous expression $G = H - TS$, where $H$ is the enthalpy and $G$ is the Gibbs free energy, such that a reaction occurs spontaneously if $\Delta G < 0$.} Then, as a whole, the protein is more likely to be unfolded than folded. This process of unfolding is called denaturation, and the fact that adding heat denatures proteins is a central reason why food tastes better when cooked!

\subsection{Machine learning using Markov random fields}\label{s:mrf}
One way of viewing the Boltzmann distribution is that it takes a set of real numbers, the energies $E(s)$ of states, and transforms them into normalized probabilities $p(s)$ for the states. Since machine learning often requires a mapping from real numbers to probabilities, many operations can be seen as a special case of the Boltzmann distribution. For instance, the softmax operation used in classification tasks takes a real vector $(a_1, \ldots, a_n)^T$ and computes a vector $(p_1, \ldots, p_n)^T$ given by
$$
p_i = \frac{e^{a_i}}{\sum_{i'=1}^n e^{a_{i'}}}.
$$
These entries are interpreted as the probabilities for each of the $n$ outputs. The softmax follows from the Boltzmann distribution from setting $a_i = -E_i / kT$. 

An area of machine learning with even clearer inspiration from physics is that of energy-based models, also called undirected graphical models or Markov random fields (MRFs). Such models were first used to investigate systems of physical particles, such as magnetic dipoles in a ferromagnetic material.\footnote{The prototypical example is the Ising model.} Within computer vision, one might imagine a binary image of $n$ by $n$ pixels, each either 0 or 1. We expect images to exhibit spatial smoothness, with neighboring pixels tending to be the same. This idea of a ``natural image prior'' can be formulated by defining an energy for the image, given as something like
$$
E(\mbox{image}) = \sum_{i\leftrightarrow j} -c(x_i - 1/2) (x_j - 1/2),
$$
with $c$ a positive parameter, $x_i$ denoting pixel values, and $i \leftrightarrow j$ denoting a sum over all pairs $\{i, j\}$ of neighboring pixels. Each pair of neighbors contributes an energy of $-c/4$ if they are the same and $c/4$ if they are different. Plugging into the Boltzmann distribution with $kT = 1$, this then defines a prior distribution over all $2^{n^2}$ binary images which encourages smoothness.

In a simple example where a few pixels from the image are corrupted, this prior can be used to solve the inference problem of filling in the missing pixels. This is known as the image inpainting problem. Markov random fields can also be used for various other computer vision tasks such as image compression, image segmentation (decomposing an image into parts), and super-resolution (generating a higher-quality, magnified image).\cite{MRF}

Within these models, the energy and partition function are generally agreed not to have actual physical meaning, but are seen as being analogous with the concepts from statistical physics.\footnote{Such naming can aid in intuition, such as the use of physics language in describing algorithms like simulated annealing, gradient descent with momentum, and Hamiltonian Monte Carlo.} The temperature dependence is often conveniently ignored because other parameters (in this case the value of $c$) can be varied instead.




\pagebreak

\section{The Story and Initial Exploration}\label{story}
Now that we've set the stage, we're ready to introduce our main act: the coconuts and islanders story.\footnote{I have chosen to preserve the original setting of the story as it was communicated to me, at the risk of magnifying the trope of painting island peoples as primitive.} An attractive feature of the story is its simplicity: not only can the Boltzmann result be visualized through a computer simulation, but almost all of the results we need can be derived without calculus. In this Section, we appropriately stay simple, using ideas from only a high school level. Then, in Section \ref{s:stats}, we'll use ideas from a first-year undergraduate probability course to make some exact calculations.




\subsection{Coconuts and islanders: the story}
Here is the coconuts and islanders story in my own words, with care to add some mathematical precision:

\begin{quote}
100 people are shipwrecked on an island, which they discover is full of palm trees and 300 coconuts. The islanders split the coconuts among themselves, 3 per person, but then decide to play the following game. Each of the islanders carries their coconuts around in a bag, and whenever two islanders run into each other, they play a match of rock-paper-scissors. The loser, assuming they still have coconuts, then gives one coconut over to the winner.

Assume that any pair of islanders is equally likely to run into each other, and that everyone has a 50\% chance of winning at rock-paper-scissors. After several days have passed, what should the distribution of coconuts over the islanders look like?
\end{quote}

\subsection{Simulation}
I wrote a Python program to simulate this story. Here is the code:\\


\begin{lstlisting}
import numpy as np
import matplotlib.pyplot as plt
N = 100  # number of islanders
C = 3  # number of coconuts per islander
T = 10**6  # 1 million timesteps

# initialize the coconut array to all C's
x = np.ones(N)*C
for t in range(T):
    # choose two islanders without replacement
    # j is the one who wins at rock-paper-scissors
    i, j = np.random.choice(N, 2)
    if x[i] > 0:
        x[i] -= 1
        x[j] += 1

plt.hist(x) # plus additional plot formatting code
\end{lstlisting}


After $10^6$ interactions (events where two people on the island meet), we plot the histogram of coconut amounts per islander. The left figure shows the result for 100 people and the right figure is a repeat of the simulation for 1,000 people and 3,000 coconuts.\\

\vspace{0.2cm}

\includegraphics[scale = 0.55]{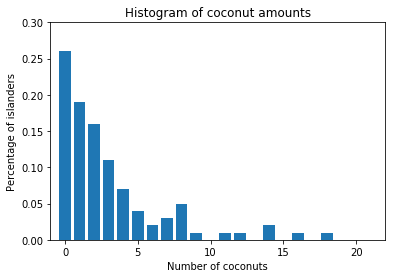}
\hspace{0.5cm}
\includegraphics[scale = 0.55]{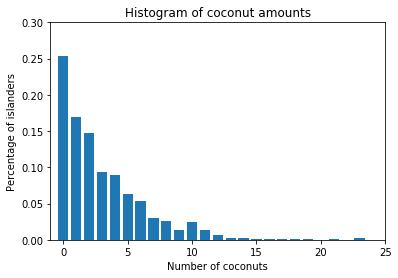}

\noindent \textbf{Figure 1}. The histogram of coconut amounts per islander over the entire island population. Left: the case of 100 islanders and 300 coconuts. Right: the case of 1,000 islanders and 3,000 coconuts.

\subsection{Initial observations: math}\label{s:initial-math}



We immediately see that the distribution in Figure 1 appears to satisfy an exponential decay. In addition, having more people on the island makes the distribution smoother while keeping the overall shape the same, a law-of-large-numbers effect.


The distribution resulting from exponential decay over the nonnegative integers is called the geometric distribution. A single parameter $p$ with $0 < p < 1$ describes the rate of decay. One can think of the geometric distribution as tossing a coin with a probability $p$ of heads, and counting the number of heads before a tails. This gives:
$$
P(X = n) = p^n(1-p).
$$

Let's for now assume that this is indeed the distribution we get, and see if we can compute $p$.\footnote{Of course, a limitation of this model is that it theoretically allows values that are arbitrarily large. In our story, there is a strict upper limit of the number of coconuts any islander can have: 300 for the first case and 3,000 for the second. However, let's deal in approximations and assume this is a fairly close model for now.} Taking the mean number of coconuts per islander to be $C$ (so $C = 3$ in our simulations), can we derive an expression for $p$ in terms of $C$? This means we want a coin that in expectation comes up with $C$ heads before a tails. We have:
$$
E(X) = \sum_{n = 0}^{\infty} n \cdot P(X = n) = C.
$$
We expand this expression and use the trick of telescoping, where we multiply both sides by $p$ and shift the placement of terms.
\begin{align*}
0\cdot (1-p) + 1\cdot p(1-p) + 2\cdot p^2(1-p) + \ldots &= C\\
0\cdot p(1-p) + 1\cdot p^2(1-p) + \ldots &= p C
\end{align*}
Subtracting the two expressions item by item gives
$$
p(1-p) + p^2(1-p) + \ldots = (1-p)C
$$
Noticing that the left side is equal to $1 - P(X = 0) = 1 - (1-p) = p$, we have
\begin{gather*}
p = (1-p)C,\\
p(1 + C) = C,\\
p = \frac{C}{1+C}.
\end{gather*}

We have our desired expression. For instance, if $C=3$, this says that a coin with probability $p=3/4$ of heads will show an average of $3$ heads before tails. Plugging in this expression for $p$, our geometric distribution is:
\begin{gather}\label{e:geometric}
P(X = n) = \frac{1}{1+C} \left( \frac{C}{1+C} \right)^n.
\end{gather}


Very well, but we can also view this as the Boltzmann distribution. Recall that the Boltzmann distribution is:
$$
p(s) = \frac{1}{Z} e^{-E(s) / kT}.
$$
For our states $s$, let's take the number of coconuts an islander could have. Then the energy $E(s)$ is naturally just this number. So
$$
P(X = n) = p(E(s) = n) = \frac{1}{Z} \left( e^{-1/kT} \right)^n.
$$
The way to identify these two expressions is by setting
\begin{gather}
e^{-1/kT} = \frac{C}{1+C},\nonumber \\
\frac{1}{kT} = \ln \left( \frac{1+C}{C} \right),\nonumber \\
kT = \left(\ln \left( 1 + \frac{1}{C} \right) \right)^{-1}.\label{e:temperature}
\end{gather}
It can then be checked that $Z$ (the normalization constant) evaluates to $1 + C$, matching (\ref{e:geometric}).

In our case, with $C = 3$, Equation (\ref{e:temperature}) evaluates to $kT = 3.48$. For $C = 100$, we have $kT = 100.50$. In fact, for large $C$ the approximation $kT \approx C + \frac{1}{2}$ is quite accurate.

Thus, we've seen that our distribution in the coconuts and islanders story looks very much like a Boltzmann distribution where the energy $E(s)$ is the number of coconuts carried by an islander, and the temperature factor $kT$ is approximately the mean number of coconuts per islander.\footnote{We'll end up verifying Equation (\ref{e:temperature}) in Section \ref{s:temperature}.}

\subsection{Initial observations: physics}\label{s:initial-physics}

So far we've only provided an argument from visually inspecting Figure 1. In Section \ref{s:stats}, we dive into rigorous mathematical arguments. But before then, let's examine the coconuts and islanders story from a physics perspective.

First of all, this simple setting illustrates a deep concept in physics: \emph{irreversibility}. Irreversibility describes processes which spontaneously operate in one direction but not the reverse, such as two liquids mixing or a gas expanding to fill its whole container. In our case, we initialized our simulation to have all islanders start out with 3 coconuts. Yet when we let the simulation run and then check in on it, it will with very high probability have reached the Boltzmann distribution. Especially with more islanders, it becomes increasingly unlikely to witness the reverse process of going from the Boltzmann distribution back to the equal initial state.

Irreversibility is closely tied with the Second Law of Thermodynamics, ``entropy always increases.'' In our case, there is only one way for the system to be perfectly ordered, whereas it turns out there are many configurations which give something looking like the Boltzmann distribution. Therefore, the system is more likely to move in a direction which allows for more configurations, or equivalently entropy. Yet given enough time, it is still possible for the system to return to its ordered state. This has led physicists and philosophers to debate whether the Second Law qualifies as a fundamental fact of nature on par with the other laws of physics.

We make one more observation on two ways of looking at our system. First is the viewpoint we've adopted up to now, where we look at how the coconuts are distributed across the island. This is called the \emph{microcanonical ensemble} in physics, and when plotted as a histogram as in Figure 1, yields \emph{Maxwell-Boltzmann statistics}. It's no coincidence that this is the same naming as the Maxwell-Boltzmann distribution for ideal gases. In fact, it's a good idea to identify our coconuts and islanders system with an ideal gas. The islanders correspond to molecules, each carrying some amount of coconuts, which represent energy. When two gas molecules meet in a collision, there is a random exchange of energy between them. Similarly, two islanders will exchange coconuts as determined by the rock-paper-scissors game. In both cases, increasing the temperature corresponds to increasing the mean number of coconuts per islander, or the mean energy per molecule.

However, there is a second perspective which focuses in on a single islander, or a single molecule. We can watch our special islander over the course of some days on the island, and track their number of coconuts over time. If we wait a long time and then plot all the observed counts on a single histogram, what will will see? It turns out we will get the same exponential decay distribution.\footnote{In Figure 1 we obtained an exponential decay curve for a single point in time, plotting all islanders. Assuming this is the true distribution, then if we overlay many points in time, we will still get an exponential decay. But this is made up of many individual islanders who are all essentially the same---in statistical terms, the individual islanders are \emph{exchangeable}. Thus, if we separate out the points coming from one islander, we are sure to also get the exponential curve from Figure 1.} In other words, the Boltzmann distribution is valid both across the islanders and for a single islander. This second viewpoint is called the \emph{canonical ensemble} and is more closely related to the protein folding and MRF examples. Just like an islander is most likely to have $0$ coconuts (the mode of the distribution in Figure 1), the protein and MRF systems tend to be found in a low-energy states (for low temperatures, at least). In the molecular gas, we get a canonical ensemble from looking a single molecule, which also prefers low-energy states. However, this is not true for the entire gas, a microcanonical ensemble, for which the total energy is fixed and cannot change.

If this is a bit confusing, there will be more about how these two viewpoints connect in Sections \ref{s:histograms} and \ref{s:ensemble}.

\pagebreak

\section{Proving the Boltzmann Result} \label{s:stats}
In this section we'll undertake a mathematical exploration into the properties of the coconuts and islanders story. There are two key concepts we'll use: a combinatorics calculation and the theoretical toolkit of Markov chains.

\subsection{Markov chains on the island}
Let's generalize our simulation to include $N$ islanders and $M$ coconuts. We want to start from the dynamics of the island game, and compute the distribution that one islander is likely to have for their amount of coconuts.

It's natural to express these dynamics in terms of a Markov chain, which you can think of as a random walk over a directed graph. Starting at one vertex in the graph, the process randomly selects an outgoing edge (weighted by its probability) and follows that edge to the next vertex.\footnote{I won't offer a full refresher of Markov chains, so it may be worth briefly reading up on them, especially when we get to the next section on stationary distributions.}

To define our Markov chain, we first need to specify the vertices in our graph, also called the state space. In order for our process to be a valid Markov chain, our states will need to include a full description of how many coconuts each islander has.

Concretely, first order the islanders $1$ to $N$ in some arbitrary fashion, e.g. in increasing order of height. Let $X_i$ be a random variable of the number of coconuts that person $i$ has. Since there are $N$ islanders and $M$ coconuts, the full state can be summarized as the tuple of numbers
$$
X_1 = n_1, X_2 = n_2, \ldots, X_N = n_N \rightarrow (n_1, n_2, \ldots, n_N)
$$
with conditions
\begin{gather*}
n_1 + n_2 + \ldots + n_N = M,\\
n_1, n_2, \ldots n_N \in \{0, 1, 2, \ldots \}.
\end{gather*}
We'll call each of these tuples a \emph{configuration}.\footnote{In physics terminology, each configuration $(n_1, n_2, \ldots, n_N)$ fully specifies the microcanonical ensemble, so they are conventionally called \emph{microstates}. See the appendix for more.}

Next, we look at our edges. Let's call two configurations $c$ and $c'$ \emph{adjacent} if it's possible to go from one to the other during a single transition step, as specified in the Python simulation. If $c$ and $c'$ are distinct and adjacent, let's call them \emph{neighbors}.\footnote{Note that if it's possible to go from $c$ to $c'$ in one step, then the reverse is also possible, so being adjacent and being neighbors are both symmetric relations.} The reason we need this second case is that some configurations can be self-adjacent, if one of the islanders in the configuration has $0$ coconuts. For instance, if two islanders with $0$ and $5$ coconuts meet, then there is a $50 \%$ chance that no exchange happens (if the islander with $5$ coconuts wins at rock-paper-scissors). If the two islanders that meet both have $0$ coconuts, then there is a $100 \%$ chance of no exchange.

For $c$ and $c'$ to be neighbors, one of the islanders must have one more coconut in $c'$ than $c$, and another islander must have one less coconut. Thus, each configuration $c$ has at most $N(N-1)$ neighbors, since we first pick two islanders with order. If a configuration $c$ has $K$ islanders with $0$ coconuts, then it only has $N(N-1) - K(N-1)$ neighbors, because $K(N-1)$ attempted transitions are invalid.

Now, we are ready to write down the transition dynamics for this Markov chain. Starting from configuration $c$, there is a $1/N(N-1)$ probability of going to configuration $c'$, for each $c'$ which is a neighbor of $c$. In the case where $K > 0$, this does not exhaust all the possible transitions. This is because $c$ is self-adjacent, with remaining probability $K(N-1)/N(N-1) = K/N$ for the chain to remain at $c$.




\subsection{Markov chain stationary distribution}\label{s:stationary}

With this description in place, we can compute a \emph{stationary distribution} for our Markov chain.\footnote{In our case, the stationary distribution describes the long-term probability distribution over configurations. In the first story we told, we started with the configuration $c = (3, 3, \ldots, 3)$. After one time step, we're guaranteed to be at an adjacent configuration. But after many time steps, the chain ``mixes'' through the main possible configurations, and converges to a fixed distribution. This stationary distribution is the same regardless of the starting configuration.
} We'll first show that the Markov chain is \emph{ergodic}, and then use the condition of \emph{detailed balance} to derive the stationary distribution over configurations.

Recall that for a Markov chain to have a stationary distribution, it must be ergodic. Proving ergodicity comes in two parts: the chain must be \emph{irreducible} and \emph{aperiodic}. Our chain is irreducible because all configurations are connected through some path of coconut exchanges. Our chain is also aperiodic because there are some self-adjacent configurations where it's possible to stay at the same configuration after a time step. Therefore, there exists a stationary distribution.

For Markov chains that satisfy detailed balance, there is a particularly easy way of calculating the stationary distribution.\footnote{Otherwise, one needs to resort to computing eigenvectors. I first learned this technique from reading an enlightening AMS Feature Column on Google's PageRank algorithm by David Austin\nocite{Austin}. The column was recommended to me by Dad and it was my first exposure to Markov chains.} Detailed balance says that if a certain distribution over configurations $\pi(c)$ satisfies
\begin{gather}\label{e:balance}
\pi(c) p(c \to c') = \pi(c') p(c' \to c)
\end{gather}
for all pairs $c$ and $c'$, then $\pi(c)$ is the stationary distribution.

In our case, consider any two configurations $c$ and $c'$. If they are the same, then the two sides of Equation (\ref{e:balance}) are already equal. If $c$ and $c'$ are not adjacent, then the probabilities $p(c \to c')$ and $p(c' \to c)$ are $0$. Finally, if $c$ and $c'$ are neighbors, then $p(c \to c') = p(c' \to c) = 1/N(N-1)$. Hence for any two neighbors $c$ and $c'$, we want $\pi(c) = \pi(c')$.

But since all configurations are connected through some path of neighbors, this means $\pi(c)$ is a constant! Following physics notation and letting $\Omega_{tot}$ be the number of configurations, we have that
$$
\boxed{\pi(c) = \frac{1}{\Omega_{tot}}, \text{ for all } c}
$$
is the stationary distribution.

This surprising observation can be summed up with the following statement: \textbf{in the limiting distribution, all configurations are equally probable.} In fact, the equivalent statement in statistical physics, ``all microstates are equally probable,'' is so important that it is called the \emph{fundamental assumption of statistical mechanics}.

\subsection{Counting configurations}

Now, we might be interested in counting $\Omega_{tot}$ to know how many configurations we really have. We do so using a combinatorics technique called ``stars and bars.'' Every configuration is a tuple of nonnegative integers $(n_1, n_2, \ldots, n_N)$ which sums to $M$. Each configuration corresponds uniquely to a cartoon drawing containing $M$ stars and $N - 1$ bars in a line. For example, in the case that $M = 7$ and $N = 3$, the configuration $c = (1, 4, 2)$ can be drawn as
$$
\star | \star \star \star \star | \star \star
$$
and the drawing
$$
\star \star \star \star \star | \star \star |
$$
can be converted to the configuration $c = (5, 2, 0)$.

Based on this correspondence, the total count $\Omega_{tot}$ is equivalent to choosing the position of $N - 1$ bars among $M + N - 1$ total symbols, which is given by:
\begin{gather}\label{e:tot}
\Omega_{tot} = \binom{M + N - 1}{N - 1}.
\end{gather}

\subsection{Tolstoy joins the party}\label{s:canonical}

We still have yet to derive the Boltzmann distribution in the coconuts and islanders story. We'll start out by focusing on the marginal distribution of coconuts for one islander, which we argued in Section \ref{s:initial-physics} should follow the same Boltzmann distribution.

The marginal distribution asks for $p(X_i = n)$, the long-term distribution of the number of coconuts that a particular islander has. Clearly the possible values of $n$ are the integers from 0 to $M$. But how is the probability mass distributed? Here, we\footnote{Taking inspiration from my Dad, who loved this quote and frequently used it in his own teaching.} follow the lead of Leo Tolstoy, who wrote in the first sentence of \emph{Anna Karenina}:
\begin{quote}
Happy families are all alike; each unhappy family is unhappy in its own way.
\end{quote}
According to this line, the reason unhappy families are so common is because there are many ways for families to go wrong. Similarly, an islander is more likely to be unhappy---have few coconuts---because there are comparably more configurations that lead to this result.\footnote{This theme once led to an spirited debate with my Dad on economic inequality, with him arguing that even perfect equality of opportunity would produce inequality, and me arguing that most societies had much more wealth concentration (and lack of mobility) than in this example.}

In the extreme case that $X_i = M$, there is only one corresponding configuration, which makes it very unlikely. On the opposite end, if $X_i$ is small, then there are many possible configurations for distributing the rest of the coconuts among $N - 1$ islanders. Since all configurations are equally probable, the way to determine likelihood is to count the number of ways.

The exact calculation of $p(X_i = n)$ is quite simple. Each setting $X_i = n$ is just a combination of the corresponding configurations, so we can write
$$
p(X_i = n) = \sum_{c \text{ with } X_i = n} \pi(c) = \sum_{c \text{ with } X_i = n} \frac{1}{\Omega_{tot}} = \frac{\Omega(X_i = n)}{\Omega_{tot}}
$$
To count the number of configurations $\Omega(X_i = n)$, we can again use the stars and bars method. If we know islander $i$ has $n$ coconuts, then there are $M - n$ coconuts left to split among the remaining $N - 1$ islanders. Each such configuration can be described by choosing $N - 2$ bars among a sequence of length $M + N - n - 2$, so
$$
\Omega(X_i = n) = \binom{M + N - n - 2}{N - 2}.
$$
Hence
\begin{gather}\label{e:canonical}
\boxed{p(X_i = n) = \frac{\Omega(X_i = N)}{\Omega_{tot}} = \frac{\binom{M + N - N - 2}{N - 2}}{\binom{M + N - 1}{N - 1}}}
\end{gather}
This is the exact marginal distribution.\footnote{The normalization condition for this distribution,
$$
\sum_{n = 0}^M p(X_i = n) = 1
$$
follows from the equation
$$
\sum_{n = 0}^M \binom{M + N - n - 2}{N - 2} = \binom{M + N - 1}{N - 1},
$$
which is a result of the so-called hockey-stick identity.}

Let's consider a particular regime where $M, N \gg 1$ and $n \ll M$.\footnote{This is common physics notation for the concept ``much greater than'' and ``much less than.'' It's not expected to be used too rigorously, and basically justifies certain approximations / throwing away terms.} We'll use a very crude approximation on Equation (\ref{e:canonical}) to see how it decreases in $n$. Expanding out the binomial coefficients, we have:
\begin{align*}
p(X_i = n) &= \frac{\binom{M + N - n - 2}{N - 2}}{\binom{M + N - 1}{N - 1}}\\
&= \frac{(M+N-n-2)!}{(N-2)!(M-n)!} \cdot \frac{(N-1)!M!}{(M+N-1)!}\\
&= \frac{(N-1) M (M-1) \cdots (M-n+1)}{(M+N-1) (M + N - 2) \cdots (M + N - n - 1)}\\
&\approx \frac{N \cdot M^n}{(M+N)^{n+1}}\\
&= \frac{N}{M+N} \left( \frac{M}{M+N} \right)^n.
\end{align*}
If we further substitute $C = M / N$ to be the mean number of coconuts per islander, as in Section \ref{s:initial-math}, we have
$$
p(X_i = n) = \frac{1}{1 + C} \left( \frac{C}{1+C} \right)^n,
$$
matching Equation (\ref{e:geometric}), a geometric or Boltzmann distribution.


\subsection{Histograms}\label{s:histograms}

We've shown that the marginal distribution for one islander is approximately geometric. But we haven't explained why Figure 1, a histogram of coconuts per islander, also yields the geometric distribution. There are two possible approaches I'm aware of, which I'll only sketch at a high level.

The first approach observes that the histogram plots the values $\{X_1, X_2, \ldots, X_N\}$ at a given point in the simulation. Each of the $X_i$ marginally follows the geometric distribution. We're not done though, because the $X_i$ are not independent of each other. In fact, we have negative correlations between each pair $X_i$ and $X_j$: if person $i$ has more coconuts, then person $j$ is expected to have fewer coconuts because the total number is fixed.

However, we argue in a heuristic fashion that as $N$ grows while $M/N$ stays constant, the marginal distributions $X_i$ stay roughly the same, while the correlations $\mbox{Corr}(X_i, X_j)$ get weaker and weaker. Thus, a joint sample of values $\{X_1, X_2, \ldots, X_N\}$ should approach the marginal distribution.\footnote{There could be a small hole in this argument, and I welcome any corrections.}

The second approach is one that I believe my Dad to have followed when he worked through the coconuts and islanders problem. We summarize the distribution using the \emph{maximum-entropy} or most likely \emph{histogram}. While each configuration $(n_1, n_2, \ldots, n_N)$ is unique and hence equally likely, each histogram is an unordered collection of values $\{n_1, n_2, \ldots, n_N\}$ and can arise from multiple configurations. Concretely, let
\begin{align*}
h_0 &= \# \{n_i = 0\},\\
h_1 &= \# \{n_i = 1\},\\
&\ldots
\end{align*}
Then one can compute the single histogram $(h_0, h_1, \ldots)$ that captures the most configurations. Details of such a computation can be found in a few textbooks\cite{Driving, Tolman}, which in practice allow for the $h_i$ to take on continuous values and apply a Lagrange multiplier method for finding the optimum under a constraint. The resulting solution follows the same Boltzmann distribution we have already seen.

\pagebreak

\section{Where Next? A Roadmap} \label{physics}

Our discussion of the coconuts and islanders story has reached a natural breathing point. Starting from a computer simulation, we've identified the key features of the coconut distribution and then provided a proof using probability and combinatorics. We can now say that the most probable distribution of coconuts across the islanders will look like a geometric or Boltzmann distribution. Furthermore, each individual islander will have a coconut amount that follows the same distribution, if sampled over many points in time (or over different random repeats of the process).

This is sufficient from a mathematical perspective. However, the coconuts and islanders story was intended to help build physical intuition. From a physics perspective, the Boltzmann distribution is concerned with real-world systems like the molecules in a gas or the configurations of a protein (Sections \ref{s:gas} and \ref{s:protein}). So there is the question of how our simple story maps onto those systems. Most pressingly, there is the question of how temperature enters into the model.

To venture further, one does need to build in the physics / chemistry side of things, most likely with the help of a textbook. For a start, I've included an appendix with a standard physics derivation of the Boltzmann distribution, adapted from Schroeder, \emph{An Introduction to Thermal Physics}. I would recommend taking a look at that derivation. Then before ending this guide, this section will aim to motivate a few last connections which are left up to the reader to explore.

\subsection{Temperature} \label{s:temperature}


In the coconuts and islanders story, all we have introduced are coconuts, which we said represent energy, and islanders, which might represent molecules in a gas. Amazingly, we can derive a temperature of this system and show that it matches up with the expression
$$
kT = \left(\ln \left( 1 + \frac{1}{C} \right) \right)^{-1},
$$
which we have already justified in Sections \ref{s:initial-math} and \ref{s:canonical}.

To start, we require the definition of thermodynamic temperature from statistical mechanics. This is actually more complicated than one might expect at first, but turns out to be incredibly powerful. Temperature is defined by a pair of equations:
\begin{gather*}
S = k \ln \Omega,\\
\frac{1}{T} = \frac{\partial S}{\partial E}.
\end{gather*}

Here $\Omega$ is a count of the number of configurations, or microstates in physics language. $S$ is the entropy, where $k$ represents Boltzmann's constant. Finally, inverse temperature $1/T$ is defined as a partial derivative of entropy with respect to energy, holding other variables constant. Using these two equations, we can write:
$$
\frac{1}{kT} = \frac{1}{k} \frac{\partial S}{\partial E} = \frac{\partial \ln \Omega}{\partial E}.
$$

On the island, there are $M$ total coconuts and $N$ islanders. Earlier in Equation (\ref{e:tot}), we counted the total number of configurations to be:
$$
\Omega_{tot} = \binom{M + N - 1}{N - 1}.
$$
So we can write down the expression $1/kT$ as
$$
\frac{1}{kT} = \frac{\partial \ln \Omega}{\partial E} = \frac{\partial}{\partial M} \Omega_{tot}(M, N) = \frac{\partial}{\partial M} \left[ \ln \binom{M + N - 1}{N - 1} \right].
$$
Using Stirling's approximation that for large $N$,
$$
\ln N! \approx N \ln N - N,
$$
we can continue to simplify as
\begin{align*}
\frac{1}{kT} &= \frac{\partial}{\partial M} \left[ \ln \left( \frac{(M+N-1)!}{(N-1)! M!} \right) \right]\\
&= \frac{\partial}{\partial M} \left[ \ln ((M+N-1)!) - \ln ((N-1)!) - \ln (M!) \right]\\
&\approx \frac{\partial}{\partial M} \left[ (M+N-1) \ln (M + N - 1) - (N - 1) \ln (N - 1) - M \ln M \right]\\
&= \ln (M + N - 1) + \frac{M+N-1}{M+N-1} - \ln(M) - \frac{M}{M}\\
&= \ln \frac{M + N - 1}{M}\\
&\approx \ln \left(1 + \frac{N}{M} \right).
\end{align*}
Recalling that earlier we set $C$ to be the mean number of coconuts per islander, or $C = M/N$, we obtain
$$
kT = \left(\ln \left( 1 + \frac{1}{C} \right) \right)^{-1},
$$
matching Equation (\ref{e:temperature}) from earlier.

\subsection{Ensembles: microcanonical, canonical, and grand canonical}\label{s:ensemble}
In the appendix, the derivation of the Boltzmann distribution that is given applies to the canonical ensemble. We commented earlier (Section \ref{s:initial-physics}) that the canonical ensemble refers to focusing on a single islander (molecule), while the microcanonical ensemble refers to looking at the whole island (gas of molecules). In physics definitions, \textbf{the microcanonical ensemble has its total energy and number of particles fixed}. Physicists always start their modeling by assuming an isolated system, and in our story, the island achieves that---ensuring that the numbers of coconuts and islanders are both conserved.

\textbf{The canonical ensemble has its number of particles fixed but can exchange energy with its surroundings.} We typically call the canonical ensemble the ``system'' and its surroundings the ``reservoir.'' We also assume that the reservoir is so large that it keeps the system at a more or less constant temperature. In our case, the system becomes a single islander, and the remaining $N-1$ islanders together constitute the reservoir. Our system can exchange energy (coconuts) with the reservoir, but the total energy of system plus reservoir is conserved.

If one reads through the derivation in the appendix, here is the key argument: if the system gives up its energy to the reservoir, the reservoir will in turn have more energy. But with more energy, there are more possible configurations for the reservoir.\footnote{This is because we assume the reservoir has a positive temperature, so $1/kT = \partial(\ln \Omega) / \partial E$ is positive.} So it's more likey that the system (islander) will be found in a low-energy (low-coconut) state. This is the essential shape of the Boltzmann distribution, and we followed this sort of reasoning in Section \ref{s:canonical}.

There is one other ensemble that shows up frequently in thermodynamics, called the grand canonical ensemble. \textbf{The grand canonical ensemble can exchange both particles and energy with its surroundings.} This is a more realistic model in studying chemical diffusion, where we might have a grand canonical ensemble on one side of a (semi-)permeable membrane. In the coconuts and islanders story, we could imagine our island is near a neighboring big island, with its own coconuts, and islanders have figured out how to travel back and forth by rowboats. Then our small island would be a grand canonical ensemble. Alternatively, a sub-area of the single island---say its north beach---can be a grand canonical ensemble, with islanders moving in and out.\footnote{The grand canonical ensemble region needs to be sufficiently small so that the other region acts like a ``reservoir,'' preserving the \emph{chemical potential $\mu$}.}

\subsection{Detailed balance and the fundamental assumption}
``In the microcanonical ensemble, all configurations are equally probable.'' This statement is called the fundamental assumption of statistical mechanics, and is a key part of the Boltzmann distribution proof. In our treatment, we proved this as a result of the stationary distribution of our Markov chain (Section \ref{s:stationary}). Yet in most undergraduate physics textbooks I encountered, there was little space taken to justify this assumption.

I would like to suggest that the statistical framework of detailed balance provides a satisfying undergraduate-level justification for the fundamental assumption.\footnote{From a pedagogical perspective, I find it unfortunate that this theme is not more present in today's textbooks (Reif's being an exception), when it was very prominent in Boltzmann's own development of statistical mechanics. Instead, authors frequently defer to a principle of insufficient reason, an argument I trace back to Tolman's influential \emph{The Principles of Statistical Mechanics} (1938). I find this line of reasoning unsatisfying, yet in Tolman's defense, Moore (2015) argues that up until recently, it remained unclear whether real-world systems actually satisfied the requirements of ergodic theory (see below).

This points to a unique feature of the coconuts and islanders story. By explicitly constructing a stochastic process, it becomes possible to prove a special case of the fundamental assumption using Markov chains, as we did in Section \ref{s:stationary}. There are a few textbooks such as Dill et al. that derive a similar distribution over nonnegative integers as ours, but their starting point is to assume equiprobable microstates.} As you might recall, detailed balance is a property of certain Markov chains, where there is a distribution $\pi(c)$ over configurations which satisfies
\begin{gather*}
\pi(c) p(c \to c') = \pi(c') p(c' \to c)
\end{gather*}
for all pairs $c$ and $c'$. Intuitively, when such a Markov chain reaches its steady state, the ``probability flow'' from $c$ to $c'$ is equal to the reverse flow from $c'$ to $c$. Markov chains which satisfy detailed balance are called \emph{reversible}, while those that don't are called \emph{irreversible}. For reversible Markov chains, if you took a video of the Markov chain and played it in reverse, the result would be indistinguishable from the forward process.

Detailed balance implies the stationary distribution is $\pi(c)$, but it does not in general imply a uniform distribution over microstates. To get a uniform distribution, one requires $p(c \to c') = p(c' \to c)$ for all pairs $c$ and $c'$. In our case, we specified a transition function, namely the rock-paper-scissors dynamic, which made this true. If we take the analogous example of a molecular gas, trying to write down the exact stochastic dynamics of all the particles would be impossible. However, we might imagine that pairs of gas particles bump into each other and exchange energy in a way that similarly satisfies this condition---if two particles of energies $E_1$ and $E_2$ meet, then the probability density that they leave with energies $E_1'$ and $E_2'$ is the same as that of starting at $E_1'$ and $E_2'$ and going to $E_1$ and $E_2$. For instance, a billiard ball model---treating atoms as tiny hard spheres---can be used to derive this, if we assume, as Boltzmann did, that the directions of the spheres prior to collision are sampled independently and uniformly at random.\footnote{This is the so-called ``molecular chaos hypothesis.'' See Reif Sections 14.2 and 14.3 for a discussion of this derivation.}



At a high level, physical systems satisfy detailed balance because of the \emph{time-reversal symmetry} of physical laws, like $F = ma$ or the Schr{\"o}dinger equation. If you take a perfect physics simulation and play it backwards, the reverse video always satisfies the same physical laws. Thus, physical systems that seem random, like molecules in a gas, always behave like a reversible Markov chain. Beautifully and mysteriously, irreversibility (the Second Law of Thermodynamics) arises from reversibility via detailed balance and the equal probability of microstates.\footnote{Reif Chapter 15 begins to discuss these properties.}

This gets us into the realm of deep philosophical issues which were first debated in the late 19th century. Boltzmann proposed his views before the discovery of quantum mechanics, yet relied heavily on the language of probability and statistics in his work. So what, if any, is the essence of randomness that underlies the Boltzmann distribution? One possible explanation is that randomness necessarily comes from underlying quantum interactions, but this turns out to be incorrect---in fact, quantum systems can show very regular behavior, such as in the quantum harmonic oscillator. A closer answer is that the randomness we describe really just represents our epistemic uncertainty, a sort of simplifying assumption that makes it possible to calculate the results we seek.

The most rigorous answer to this question comes from the branches of mathematics called chaos theory and ergodic theory.\footnote{Fun fact: the word ``ergodic'' was coined by no less than Boltzmann himself!} The layperson's definition of chaos theory is that small changes in initial conditions can lead to magnified differences down the road---the so-called ``butterfly effect.'' For systems that exhibit these effects, no amount of precision is enough to completely determine future trajectories, thus yielding the appearance of randomness. Ergodic theory further provides a language for speaking of such systems in terms of probability, essentially justifying the Markov chain view that we have taken throughout this piece. From an ergodic theory viewpoint, the fundamental assumption is called the \emph{ergodic hypothesis}.\footnote{In this perspective, the time-reversal symmetry of physical laws enters via the existence of a Hamiltonian, which implies a result called Liouville's Theorem that is necessary (but not sufficient) for systems to be ergodic.}

In summary, the developments of chaos theory and ergodic theory are necessary to provide a formal footing and exclude special edge cases, but overall they have vindicated Boltzmann's approach. They imply that complicated deterministic systems can often be rewritten with a simpler random model. Thus, at an introductory level, detailed balance is a valid way to justify that ``all microstates are equally probable.''\footnote{The situation is actually more subtle for quantum mechanics, and is still not completely resolved! See the Wikipedia articles on many body localization and the eigenstate thermalization hypothesis for a start into this active area of research.}

\subsection{Extensions}
Hopefully this piece inspires other questions and extensions, ultimately leading the reader to a more intuitive framework for thinking about the Boltzmann distribution. Here are some follow-up areas I've thought of but have lacked the time or expertise to pursue in detail.



\begin{itemize}
\item As a distribution over the nonnegative integers, the Boltzmann distribution turns out to be identical to the geometric distribution. One way of seeing that the distribution in the story is really Boltzmann is to impose a bag size limit of 10 coconuts per islander. If two islanders play rock-paper-scissors and the winner has a full bag, no exchange takes place. The resulting distribution will be Boltzmann over the integers from 0 to 10. Additionally, if the mean number of coconuts per islander is greater than 5, we actually enter the regime of \emph{negative temperature}, which shows up in physical systems as well.

\item Instead of a discrete state space, one can extend the simulation to a continuous state space. For instance, all the islanders start out with 3 liters of rainwater. Whenever two islanders meet, they simulate a uniform real number from -1 to 1, which is how much water gets exchanged from the taller islander to the shorter islander, assuming both islanders can maintain a nonnegative amount of water. A variety of transition dynamics can be used (so the simulation could be from the standard normal as well), as long as detailed balance is still satisfied.

\item One could try to make a realistic simulation in three dimensions that recovers the Maxwell-Boltzmann distribution, either over a discretized or a continuous space, with each molecule's velocity $(v_x, v_y, v_z)$ taking integer or real-valued coordinates respectively. Spatial locality would be another aspect to model, so that close particles are more likely to collide.

\item Perhaps it is possible to modify the transition dynamics in the coconuts and islanders problem to simulate Fermi-Dirac or Bose-Einstein statistics.

\item From an economic modeling perspective, one could allow the outcome of rock-paper-scissors to favor certain fixed islanders (systemic advantage), wealthy islanders (rich-get-richer), or poor islanders (affirmative action).\footnote{See Vi Hart and Nicky Case's ``Parable of the Polygons''\nocite{Polygon} for a similar societal simulation.}


\item From a computational perspective, one might want to investigate whether the ``coconuts and islanders'' Markov chain described here might inspire new ways of simulating from energy-based statistical models. One observation is that the pairwise islander interactions naturally leads to parallelization.



\end{itemize}


\pagebreak

\section{Conclusion}\label{conclusion}

\subsection{A tribute to Shoucheng Zhang (1963 - 2018)}
My father, Shoucheng Zhang, was a theoretical physics professor at Stanford until he passed away in 2018. He worked in the area of condensed matter physics, which is concerned with novel states of matter that arise from collections of quantum-interacting particles. A key theme in condensed matter physics is the emergence of macroscopic order out of microscopic disorder, of which the Boltzmann distribution is one example. Dad started his career working on models of superconductivity, and later played a key role in the discovery and development of a new state of matter called topological insulators.

As mentioned earlier, Dad told me and my sister the coconuts and islanders story as one of his many science illustrations.\footnote{When I asked him about the origin of the story a year or two ago, he said he may have heard it from a colleague, in which case I'll gladly add an attribution.} I think his own understanding of the story was less bogged down by mathematical detail, and probably involved a maximum entropy argument with the starting assumption (or intuition) that all microstates are equally probable. This was one feature of my Dad's approach to understanding: usually he cared most about the crux mathematical element in a calculation, while other mathematical formalism was seen as a bit of a distraction from this elegance. This made him a great teacher of physics to me, as I knew I would receive the most distilled explanation.

Dad greatly enjoyed teaching, with a highlight being a Stanford freshman introductory seminar (``introsem'') he designed around dimensional analysis and back-of-the-envelope physics. He believed teaching to a young audience was one of the best ways to consolidate his own understanding, as it required simple and accessible explanations. Many of his graduate students have also described a mentorship style of encouraging basic and intuitive explanations.

Dad often used the motto ``simplicity and universality'' to describe his quest as a theoretical physicist. I hope that my retelling of the coconuts and islanders story meets these criteria. I can't imagine a simpler story than this one to illustrate statistical mechanics, and in my analysis I've sought to make maximum use of elementary tools like computer simulation and stars and bars counting. Yet the story serves as a case study of the Boltzmann distribution, which underpins all of chemistry and thus much of what we experience every day. Like James Joyce's Dublin, pedagogical examples such as these are worth revisiting over and over, ``because if I can get to the heart of Dublin I can get to the heart of all the cities of the world. In the particular is contained the universal.''





\subsection{Acknowledgements and further reading}
A full list of references and recommended reading is included at the end of this writeup.

Two standard thermal physics textbooks are Daniel V. Schroeder's \emph{Introduction to Thermal Physics}, which I used as an undergraduate, and Frederick Reif's \emph{Fundamentals of Statistical and Thermal Physics}, which is a bit more advanced and which Dad used in his teaching\nocite{Schroeder, Rief}. For reading about the Boltzmann distribution and its applications, my go-to recommendation would be \emph{Molecular Driving Forces}\nocite{Driving} by Dill, Bromberg, and Stitger. Although the textbook focuses on applications of thermodynamics in chemistry, I've found it unmatched in terms of clarity and illustrative examples, all while covering the same results found in a physics textbook. My protein folding example was adapted from this book.


Werner Krauth's excellent article ``Introduction To Monte Carlo Algorithms''\nocite{Krauth} inspired my writing style and reminded me to keep things simple, readable, and fun. I also enjoyed reading parts of Boltzmann's \emph{Lectures on Gas Theory}\nocite{Boltzmann}, as well as the Wikipedia articles on Boltzmann's H-theorem, microscopic reversibility, detailed balance, Loschmidt's paradox, and the eigenstate thermalization hypothesis.

For more on the mathematical underpinnings of the fundamental assumption of statistical mechanics in both classical and quantum systems, I've found a few excellent but challenging review articles in the area. Oliveira and Werlang (2007)\nocite{Oliveira} and Moore (2015)\nocite{Moore} give an overview of the classical case (ergodic theory), while D'Alessio et al. (2016)\nocite{DAlessio} focuses on quantum chaos.


For more discussion on energy-based statistical models, I would recommend chapter 8 of Christopher Bishop's \emph{Pattern Recognition and Machine Learning}\nocite{Bishop}, on which my MRF example was based. Alternatively, see David MacKay's \emph{Information Theory, Inference, and Learning Algorithms}\footnote{Dad introduced me to MacKay's textbook sometime when I was in college. My first time trying to learn variational inference was when he had downloaded a lecture by MacKay and we sat in front of our home TV watching together. I'm sure MacKay's physics background was a big inspiration for Dad's foray into machine learning, and he was proud of having published a paper in the field: \url{https://www.pnas.org/content/115/28/E6411}.}, which is freely available online\nocite{MacKay}. Chapters 31 and 43 cover ``Ising Models'' and ``Boltzmann Machines'' respectively.

Thanks to Jonty Rougier for pointing me to the tufte-handout TeX template, which I'm grateful to many contributors, not least Edward Tufte, for designing. Thanks to Andriy Mnih, Arpon Raksit, Rahul Dalal, Stephen Mackereth, and Simon Lieu for providing comments and corrections.


\newpage

\section{Appendix: Deriving the Boltzmann Distribution in General} 
In this appendix, I'll review Schroeder's derivation of the Boltzmann distribution.\cite{Schroeder} The exposition is not original, but rather, I wanted an available reference for the standard physics-based presentation of these concepts. (Another approach derives Maxwell-Boltzmann statistics via a maximum entropy calculation; this was alluded to in Section \ref{s:histograms} and details can be found in Tolman\cite{Tolman} and Dill et al.\cite{Driving})

\subsection{Preliminaries}
Before deriving the Boltzmann distribution, let me first sketch out the important results Schroeder covers in chapters 1-3.

\textbf{Energy}. Energy is just some quantity that is always conserved (Schroeder p. 17).

\textbf{Thermal contact}. A setup where the parts of a system have been allowed to freely exchange energy with each other. (My definition.)

\textbf{Thermal equilibrium}. A setup of thermal contact that has come to a steady state. (My definition, Schroeder's is on p. 2.)


\textbf{Boltzmann constant}. $k = 1.381 \times 10^{-23}$ J/K (Joules per Kelvin, Schroeder p. 7).

\textbf{Microstate / macrostate}. Schroeder is not rigorous on this, explaining that a microstate specifies the outcome of each individual particle in a system, while the macrostate describes the state more generally (Schroeder p. 50).

\textbf{Multiplicity}. Denoted $\Omega$, this counts the number of microstates that lead to a given macrostate (Schroeder p. 50).

\textbf{Fundamental assumption of statistical mechanics}. ``In an isolated system in thermal equilibrium, all accessible microstates are equally probable'' (Schroeder p. 57). Schroeder does not offer a proof, though he does hint at the principle of detailed balance as a way of justifying this.

\textbf{Entropy}. The most general definition of entropy in thermodynamics is:
$$
S \equiv -k \sum_i p_i \ln p_i,
$$
where $i$ is an index over the different outcomes of a system, and $p_i$ are the probabilities of the individual outcomes. This is equivalent to the Shannon entropy / information up to a proportionality constant.

However, if we are interested in the entropy of a macrostate, then $i$ becomes an index over the microstates and $p_i = 1 / \Omega$ for all $i$, leading to the simpler formula (Schroeder p. 75)
$$
S = k \ln \Omega.
$$

\textbf{Second law of thermodynamics}. For systems with large numbers of particles, the majority of microstates often will be concentrated in a small number of macrostates. Due to the fundamental assumption, these macrostates are the most likely, and a system that starts in a low multiplicity macrostate will tend towards high multiplicity. Because of the entropy formula, this is equivalently stated as ``entropy tends to increase'' (Schroeder p. 59, 76).

\textbf{Temperature}. The usual definition is
$$
\frac{1}{T} \equiv \frac{\partial S}{\partial E}.
$$
The partial derivative emphasizes that other quantities, such as the volume and number of particles in the system, are held fixed (Schroeder p. 88).

\textbf{Corollaries of the definition of temperature}. Consider two systems $A$ and $B$ which are in thermal contact with each other. Due to the second law, the overall entropy $S_A + S_B$ will tend to increase. One can show based on the definition of temperature that if $T_A > T_B$, energy will be inclined to flow from $A$ to $B$, and vice versa if $T_A < T_B$. At thermal equilibrium this process comes to a halt, so that $T_A = T_B$, which is what we would expect (Schroeder p. 85-88).

\subsection{Main argument}
The condition for the Boltzmann result is that our system of interest is at a fixed temperature $T$. More formally, we consider a system that ``is in thermal equilibrium with a `reservoir' at a specified temperature'' (Schroeder p. 220). This gives us an operational definition of fixing the temperature, and also provides a model for working through the calculations. According to Schroeder, ``a reservoir in thermodynamics is anything that's so large that its temperature doesn't change noticeably when [energy] enters or leaves'' (p. 122), a definition which is good enough for our purposes.

Now we consider the combined system of our area of interest $A$ and the reservoir $R$, which we can assume is isolated from the rest of the universe. By the conservation of energy, we know that $E_A + E_R = E_{tot}$ is a fixed quantity. By the fundamental assumption of statistical mechanics, all accessible microstates of the combined system are equally probable, so the probability of a particular macrostate is proportional to $\Omega_A \Omega_R$ for that configuration.

Since our macrostates of interest are the states of $A$, let us consider two possible states $s_1$ and $s_2$. For each of these, $\Omega_A = 1$. We allow $A$ and $R$ to exchange energy, but nothing else, so we would expect $\Omega_R$ to be simply a function of $E_R$. We write
$$
\Omega_R(s) = \Omega_R(E_R(s)) = \Omega_R(E_{tot} - E_A(s)).
$$
So if we consider the ratio of probabilities of our two states, we have
$$
\frac{p(s_1)}{p(s_2)} = \frac{\Omega_A(s_1) \Omega_R(s_1)}{\Omega_A(s_2) \Omega_R(s_2)} = \frac{\Omega_R(E_R(s_1))}{\Omega_R(E_R(s_2))}.
$$
The one thing we know about the reservoir is its temperature:
$$
\frac{1}{T} = \frac{\partial S_R}{\partial E_R}.
$$
We can integrate this expression between $E_R(s_2)$ and $E_R(s_1)$, and assuming the energy difference is small compared to the capacity of the reservoir, and that other thermodynamic variables are held fixed, we have
\begin{align*}
S_R(E_R(s_1)) - S_R(E_R(s_2)) &= \int_{E_R(s_2)}^{E_R(s_1)} \left(\frac{\partial S_R}{\partial E_R} \right) dE_R\\
&= \int_{E_R(s_2)}^{E_R(s_1)} \frac{dE_R}{T}\\
&= \frac{E_R(s_1) - E_R(s_2)}{T}.
\end{align*}
Performing some rearrangements,
\begin{gather*}
S_R(E_R(s_1)) - S_R(E_R(s_2)) = \frac{E_R(s_1) - E_R(s_2)}{T},\\
k \ln \Omega_R(E_R(s_1)) - k \ln \Omega_R(E_R(s_2)) = \frac{[E_{tot} - E_A(s_1)] - [E_{tot} - E_A(s_2)]}{T},\\
\ln \left( \frac{\Omega_R(E_R(s_1))}{\Omega_R(E_R(s_2))} \right) = - \frac{E_A(s_1) - E_A(s_2)}{kT}.
\end{gather*}
Hence,
$$
\frac{p(s_1)}{p(s_2)} = \frac{\Omega_R(E_R(s_1))}{\Omega_R(E_R(s_2))} = \exp \left( - \frac{E_A(s_1) - E_A(s_2)}{kT} \right).
$$
Since this holds true for any pair of states, we deduce that
$$
p(s) \propto e^{-E(s) / kT},
$$
as desired.

This is the main argument Schroeder provides. Note that the integration was assumed to be by an overall energy difference $E_R(s_1) - E_R(s_2)$ that did not significantly affect the temperature of the reservoir. When considered over extreme energy differences that begin to drastically change the energy in the reservoir, the Boltzmann result might break down.\footnote{Stated in other terms, the Boltzmann distribution is simply the result of making a locally linear approximation to the reservoir's entropy as a function of energy. The slope one obtains is $1/T$. If the local slope was 0 (infinite temperature), then the correct local approximation would instead be of second order, taking the form $\exp(-E(s)^2c)$ for some constant $c$.}

\pagebreak

\phantomsection
\addcontentsline{toc}{section}{References}
\bibliography{boltzmann-bib}
\bibliographystyle{plainnat}

\end{document}